\documentclass[seceq]{ptptex}

\usepackage{graphicx}
\usepackage{wrapft}
\usepackage{multirow}
\newcommand{\reaction}{(p,d)}
\newcommand{\etapC}{{}^{11}\mathrm{C} \otimes {\eta'}}
\newcommand{\Cerenkov}{{\rm \check{C}erenkov}}

\title{
Feasibility Study of Observing $\eta'$ Mesic Nuclei with $\reaction$ Reaction%
}

\author{
Kenta \textsc{Itahashi}$^1$, %
Hiroyuki \textsc{Fujioka}$^{2,}$\footnote{E-mail: fujioka@scphys.kyoto-u.ac.jp}, 
Hans \textsc{Geissel}$^3$, 
Ryugo S. \textsc{Hayano}$^4$,\\
Satoru \textsc{Hirenzaki}$^5$, 
Satoshi \textsc{Itoh}$^4$, 
Daisuke \textsc{Jido}$^{6,7}$,
Volker \textsc{Metag}$^8$,\\
Hideko \textsc{Nagahiro}$^5$,
Mariana \textsc{Nanova}$^8$,
Takahiro \textsc{Nishi}$^4$,\\
Kota \textsc{Okochi}$^4$,
Haruhiko \textsc{Outa}$^1$,
Ken \textsc{Suzuki}$^9$,
Yoshiki K. \textsc{Tanaka}$^4$\\
and Helmut \textsc{Weick}$^3$
}

\inst{
$^1$Nishina Center for Accelerator-Based Science, RIKEN, Saitama 351-0198, Japan\\
$^2$Department of Physics, Kyoto University, Kyoto, 606-8502, Japan\\
$^3$GSI Helmholtzzentrum f\"{u}r Schwerionenforschung GmbH, D-64291 Darmstadt, Germany\\
$^4$Department of Physics, The University of Tokyo, Tokyo, 113-0033, Japan\\
$^5$Department of Physics, Nara Women's University, Nara 630-8506, Japan\\
$^6$Yukawa Institute for Theoretical Physics, Kyoto University, Kyoto, 606-8502, Japan\\
$^7$J-PARC Branch, KEK Theory Center, Institute of Particle and Nuclear Studies, 
High Energy Accelerator Research Organization (KEK), Ibaraki, 319-1106, Japan
$^8$II. Physikalisches Institut, Universit\"{a}t Gie{\ss}en, D-35392 Gie{\ss}en, Germany\\
$^9$Stefan Meyer Institut f\"ur subatomare Physik, 1090 Vienna, Austria
}

\abst{
A novel method is proposed to measure $\eta'$(958) meson bound states
in $^{11}{\rm C}$ nuclei by missing mass spectroscopy 
of the $^{12}{\rm C}\reaction$ reaction near the $\eta'$ production threshold.
It is shown that peak structures will be observed experimentally
in an inclusive measurement
in case that the in-medium $\eta'$ mass reduction is sufficiently large
and that the decay width of $\eta'$ mesic states is narrow enough.
Such a measurement will be feasible with
the intense proton beam supplied by the SIS synchrotron at GSI 
combined with the good energy resolution of the fragment separator FRS.

}

\begin{document}

\maketitle
\section{Introduction}
The U$_{A}(1)$ problem~\cite{Weinberg:1975ui} has attracted continuous 
attention for a long time as a fundamental question on the low-energy 
spectrum and dynamics of the pseudoscalar mesons in QCD. 
According to the symmetry pattern of the quark sector in QCD naively,
the $\eta^{\prime}$ meson would be one of the Nambu-Goldstone bosons 
associated with the spontaneous breakdown of the U(3)$_{L}\times$U(3)$_{R}$ 
chiral symmetry to the U$_{V}(3)$ flavor symmetry.  In the real world, however, 
gluon dynamics plays an important role,
and the $\eta^{\prime}$ meson acquires 
its peculiarly larger mass than those of the other pseudoscalar mesons, 
$\pi$, $K$, and $\eta$ through the quantum anomaly effect of non-perturbative 
gluon dynamics~\cite{Witten:1979vv,Veneziano:1979ec} 
which induces the non-trivial vacuum structure of QCD~\cite{'tHooft:1976up}.
The mass generation of the $\eta^{\prime}$ meson is a result 
of the interplay of quark symmetry and gluon dynamics.
In the instanton picture,
a rapid decrease of the effects of instantons in finite energy 
density hadronic matter induces a reduction of the $\eta^{\prime}$
mass~\cite{Pisarski:1983ms,Kapusta:1995ww}.
It is also known that the U$_{A}(1)$ anomaly affects 
the flavor singlet $\eta^{\prime}$ mass only through
chiral symmetry breaking in the light flavors~\cite{Lee:1996zy,Jido:2011aa}.
In spite of this qualitative theoretical understanding of the mechanism of the 
$\eta^{\prime}$ mass generation, we 
have not understood the $\eta^{\prime}$ mass generation quantitatively  
and our present knowledge has not reached a satisfactory level yet. 

In-medium properties of the $\eta^{\prime}$ meson provide us with 
essential information to understand the U$_{A}(1)$ anomaly, and the density 
and/or temperature dependence of the anomaly effect on the $\eta^{\prime}$ 
mass may clarify the mass generation mechanism.
In nuclear 
matter partial restoration of chiral symmetry takes place with an order of 30\%
reduction of the quark condensate~\cite{Suzuki:2002ae,Kolomeitsev:2002gc,Jido:2008bk}.
The effect of the chiral symmetry breaking on the $\eta^{\prime}$ mass could 
decrease and one could expect a large mass reduction in nuclear matter. 
One of the efficient ways to observe in-medium modification of the meson properties 
is spectroscopy of meson-nucleus bound systems like deeply bound
pionic atoms. In the meson-nucleus bound system, because it is guaranteed that the meson 
inhabits the nucleus, it is unnecessary to remove in-vacuum contributions 
from the spectrum. The fact that the bound states have definite quantum numbers
is favorable to extract fundamental quantities, since 
detailed spectroscopy enables us to investigate selectively the contents of the 
in-medium meson self-energy~\cite{Itahashi:1999qb}.
Since so far there have been no observations of $\eta^{\prime}$ mesons bound 
in nuclei, it is extremely desirable to search for experimental signals of 
$\eta^{\prime}$ bound states in nuclei as a first step of detailed investigations
of in-medium $\eta^{\prime}$ meson properties.

Recently it has been reported that a strong reduction of the $\eta^{\prime}$ 
mass, at least 200 MeV, is necessary to explain the two-pion correlation 
in Au+Au collisions at RHIC~\cite{Csorgo:2009pa}. 
On the other hand, a low-energy $\eta'$ production experiment with $pp$ collisions
has suggested a relatively smaller 
scattering length of the $s$-wave $\eta^{\prime}$-proton interaction, 
$|\mathrm{Re}\ a_{\eta^{\prime}  p}| <0.8\ \mathrm{fm}$~\cite{Moskal} and  $| a_{\eta^{\prime} p}| \sim 0.1\ \mathrm{fm}$~\cite{Moskal:2000pu}, 
which corresponds to a mass reduction at nuclear saturation density
from several to tens MeV estimated within the linear density approximation. 
Transparency ratios for $\eta'$ mesons have been measured for different nuclei
by the CBELSA/TAPS collaboration~\cite{Nanova_PLB, Nanova_PPNP}.
An absorption width of the $\eta'$ meson at saturation density
as small as 15--25 MeV has been found.
Within the experimental uncertainties
this width seems to be almost independent of the $\eta'$ momentum.
Theoretically, the formation spectra of the $\eta'$ mesic nuclei have been calculated first in Ref.~\citen{Nagahiro:2004qz}. Nambu--Jona-Lasinio model calculations suggested 150--250 MeV 
mass reduction of the $\eta^{\prime}$ meson at saturation 
density~\cite{Costa:2002gk,Nagahiro:2006dr}. 
We mention here a recent theoretical work~\cite{Nagahiro11} suggesting that 
the $\eta^{\prime}$-nucleus optical potential evaluated based on an effective 
Lagrangian for the $\eta^{\prime}$ nucleon scattering~\cite{OsetRamos}
has a significantly larger real part than the imaginary part as expected in Ref.~\citen{Jido:2011aa}.  
However, at the same time, the effective Lagrangian approach~\cite{OsetRamos} shows
that the existence of the strong attraction is not consistent with the latest data of the small 
scattering length \cite{Moskal:2000pu} and that,  because of the lack of the experimental 
information at present, even the sign of the real part of the potential cannot be determined.  

In the present work, our argument is based on the fact that
the anomaly effect can contribute to the $\eta^{\prime}$ mass only
in the presence of the spontaneous and/or explicit breaking of chiral symmetry~\cite{Lee:1996zy,Jido:2011aa}.
This is because the chiral singlet gluon current cannot couple to the chiral 
pseudoscalar mesonic state without chiral symmetry breaking. 
Thus, even if the density dependence of
the U$_{A}$(1) anomaly effect is moderate, a relatively
large mass reduction of the $\eta^{\prime}$ meson is expected 
at nuclear density due to the partial restoration of chiral symmetry.
For instance,
assuming a 30\% reduction of the quark condensate in the nuclear medium
and that the $\eta'$-$\eta$ mass difference 
depends linearly on the quark condensate, one could expect an attraction
of the order of 100 MeV for the $\eta^{\prime}$ meson coming from partial restoration 
of chiral symmetry in nuclear medium. 
This $\eta^{\prime}$ mass reduction mechanism 
has a unique feature.~\cite{Jido:2011aa}
In usual meson-nuclear systems, attractive interactions induced by 
many-body effects are unavoidably accompanied by comparably large 
absorptions. This is because attractive interaction and absorption processes
originate from the same hadronic many-body effects. This implies that  
the bound states have an absorption width comparable to the level spacing. 
In the present case, however, since the suppression of the U$_{A}$(1) 
anomaly effect in the nuclear medium induces the attractive interaction to the 
in-medium $\eta^{\prime}$ meson, 
the influence acts selectively on the $\eta^{\prime}$ meson and, thus, 
it does not induce inelastic transitions of the $\eta^{\prime}$ meson into 
lighter mesons, 
although other many-body effects can introduce nuclear absorptions 
of the $\eta^{\prime}$ meson.
Consequently 
the $\eta^{\prime}$ meson bound state may have a larger binding
energy with a smaller width.~\cite{Jido:2011aa}

This paper is organized as follows.
In Section 2, formation spectra for $\eta'$ mesic nuclei by the $\reaction$ reaction
on $^{12}\mathrm{C}$ are discussed.
Section 3 introduces the experimental concept by use of the $\reaction$ reaction.
The result of a simulation of the missing mass spectrum will be shown in Section 4.
Finally, Section 5 is devoted to the summary.

\section{Theoretical Calculation of $\reaction$ Spectra}
\subsection{The $(p,d)$ reaction for $\eta'$ mesic nucleus formation}

\begin{wrapfigure}{1}{6.6cm}
\includegraphics[width=6.0cm]{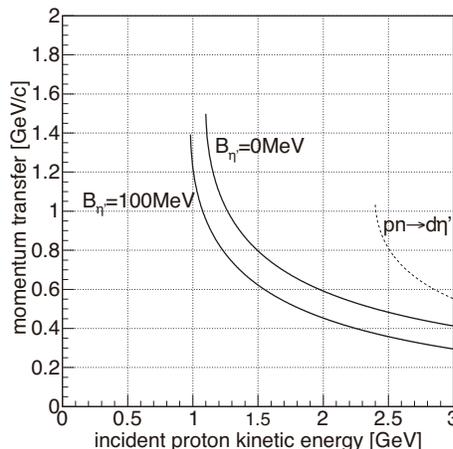}
\caption{Momentum transfer of the $^{12}\mathrm{C}\reaction\etapC$ reaction as a function of 
incident proton
kinetic energy. The solid lines correspond to the $\eta'$ binding
energy ($B_{\eta'}$) of $0\,\mathrm{MeV}$ and $100\,\mathrm{MeV}$. For
a reference, the momentum transfer for the elementary process is also
shown by the dashed line.}\label{momentum_transfer}
\end{wrapfigure}

We propose to investigate the $\reaction$ reaction on a $^{12}\mathrm{C}$
target in order to produce $\eta'$ mesic nuclei,
i.e.~$\etapC$. In this reaction, the initial proton picks up a neutron in the 
nuclear target and simultaneously an $\eta^{\prime}$ meson is created inside
the residual nucleus.
The created $\eta^{\prime}$ mesons will form bound states in the  
nuclei having a neutron hole state. Observing the emitted deuteron 
in the forward direction, we obtain missing-mass spectra for the $\eta^{\prime}$
mesic nucleus formation. The shape of the formation spectrum is determined 
by convolutions of the nucleon hole and $\eta^{\prime}$ bound state wavefunctions. 

The momentum transfer for this reaction, as a function of the incident kinetic energy,
is shown in Fig.~\ref{momentum_transfer}.
It should be noted that the recoil-free condition cannot be satisfied for
any kinetic energy in this reaction.
A larger momentum transfer compared to the ($\gamma$,$p$)
and ($\pi$,$N$) reactions~\cite{Nagahiro05, Nagahiro10}
is, however, acceptable or even desirable when taking into account the following two aspects.
\begin{itemize}
\item From an experimental point of view,
 it is difficult to detect ejectiles with the magnetic rigidity 
 (the momentum divided by the charge) too close to that of
the projectile. 
In this sense, large momentum transfer
will allow us to utilize a high intensity beam,
which may compensate a small formation probability of $\eta'$ mesic nuclei
due to large momentum transfer.
\item
In general, kinematics with small recoil momentum is preferred because
substitutional configurations of the nucleon hole and the $\eta'$ meson 
are largely populated.
On the contrary, large momentum transfer will induce
an $\eta'$ meson with large angular momentum,
and hence the strength of an excited state near the $\eta'$ production threshold
is enhanced, as explained in the next subsection.
Then, the first aim of the unprecedented spectroscopy
can be taken at the observation of a single narrow peak,
expected to be distributed near the threshold.

\end{itemize}
Thereby, the incident kinetic energy is determined
to be $2.50\,\mathrm{GeV}$,
slightly above the threshold for the elementary process ($pn\to d\eta'$),
which is around $2.40\,\mathrm{GeV}$.
While there is no direct experimental information on its cross section,
we evaluated the differential cross section at the forward angle
as $\sim 30\,\mathrm{\mu b/sr}$, 
which will be used for a theoretical calculation
of the formation spectrum.
The deduction will be described in the Appendix.

According to Ref.~\citen{Jido:2011aa}, light nuclear targets, $A \sim 10$ to $20$, 
are preferable to observe clearer structure of the $\eta^{\prime}$ bound states.%
\footnote{As shown in Ref.~\citen{Nagahiro11} for weaker attraction cases
heavier nuclei might be better for observation of bound states.}
In heavier nuclear targets, the residual nuclei have more neutron hole states.
The peaks in the formation spectra come from many possible combinations 
of the $\eta^{\prime}$ bound states and the neutron hole states. 
This makes the spectrum more complicated 
and the peak structure gets less prominent. In addition, since the strength 
of the $\eta^{\prime}$ attraction in nuclei depends on the nuclear density 
and not on the number of the nucleons,
unlike for pionic atoms, heavier nuclear targets do not help to provide stronger attraction.
Nevertheless, since the spatial size of the $\eta^{\prime}$ potential depends on 
the nucleus and since the level spacing of the bound states can be different in each 
nuclear target, observing the $\eta^{\prime}$ mesic nucleus spectra with 
different nuclear targets enables us to investigate the detailed structure of the 
$\eta^{\prime}$ potential. 

\subsection{Theoretical calculation of formation spectra}

We evaluate the formation rate of the signal process 
$p + A \rightarrow d + (A-1)\otimes{\eta'}$
in the $\reaction$
reaction by using 
the Green's function method\cite{NPA435}, in which the reaction cross section is
assumed to be separated into the nuclear response function $S(E)$ and
the elementary cross section of the $pn\rightarrow d\eta'$ process within
the impulse approximation\footnote{In this formulation, we do not include the kinematical correction factor
reported in Refs.~\citen{Dover,Koike08,Ikeno11} to take into account 
the nuclear recoil effects correctly.  This effect tends to enhance the
spectrum in $\eta'$ bound energy region slightly 
compared to that in quasi elastic region, and reduces the size
of the whole spectrum to about 40\%.}:
\begin{equation}
 \left(\frac{d^2\sigma}{d\Omega dE}
\right)_{A(p,d)(A-1)\otimes\eta'}  =
\left(\frac{d\sigma}{d\Omega}
\right)^{\rm lab}_{n(p,d)\eta'} \times S(E)\ ,
\label{eq:IA}
\end{equation}
where the nuclear response function $S(E)$ is given in terms of the
in-medium Green's function.  The detailed expressions are given in
Ref.~\citen{Nagahiro:2008rj}.
The elementary cross section $\left(\frac{d\sigma}{d\Omega}
\right)^{\rm lab}_{n(p,d)\eta'} $ is estimated to be 30 $\mu$b/sr
as described in the Appendix.
We estimate the flux loss of the injected proton and the ejected
deuteron owing to the elastic and quasielastic scattering and/or
absorption processes in the target and daughter nuclei by using the
eikonal approximation~\cite{Nagahiro:2008rj}.

For the $\eta'$-nucleus optical potential, we simply assume an
empirical form as,
\begin{equation}
 V_{\eta'}=(V_0+iW_0)\frac{\rho(r)}{\rho_0},
\label{eq:V}
\end{equation}
with the nuclear density distribution $\rho(r)$ and the normal
saturation density $\rho_0=0.17$ fm$^{-3}$.  The mass term in the
Klein-Gordon equation for the $\eta'$ meson at finite density can be
written as
\begin{equation}
 m_{\eta'}^2 \rightarrow m_{\eta'}^2(\rho) = (m_{\eta'}+\Delta
  m(\rho))^2
\sim m_{\eta'}^2+2m_{\eta'}\Delta m_{\eta'}(\rho),
\end{equation}
where $m_{\eta'}$ is the mass of the $\eta'$ meson in vacuum and
$m_{\eta'}(\rho)$ the mass at finite density $\rho$.  The mass shift
$\Delta m_{\eta'}(\rho)$ is defined as $\Delta
m_{\eta'}(\rho)=m_{\eta'}(\rho)-m_{\eta'}$.  Thus, we can interpret the
mass shift $\Delta m_{\eta'}(\rho)$ as the strength of the real part of
the optical potential
\begin{equation}
 V_0=\Delta m_{\eta'}(\rho_0)\ ,
\end{equation}
using the mass shift at the normal saturation density $\rho_0$.  Here we
assume the nuclear density distribution $\rho(r)$ to be of an empirical
Woods-Saxon form as
\begin{equation}
 \rho(r)=\frac{\rho_N}{1+\exp(\frac{r-R}{a})},
\end{equation}
where $R=1.18 A^{\frac{1}{3}}-0.48$ fm, $a=0.5$ fm with the nuclear
mass number $A$, and $\rho_N$ a normalization constant defined so that
$\int d^3r\rho(r)=A$.  We obtain the in-medium Green's function by solving 
the Klein-Gordon equation with the optical potential $V_{\eta'}$ in
Eq.~(\ref{eq:V}) with the appropriate boundary condition and use it to
evaluate the nuclear response function 
$S(E)$ in Eq.~(\ref{eq:IA}).

\begin{figure}[t]
\begin{center}
\includegraphics[angle=0, width=0.97\textwidth]{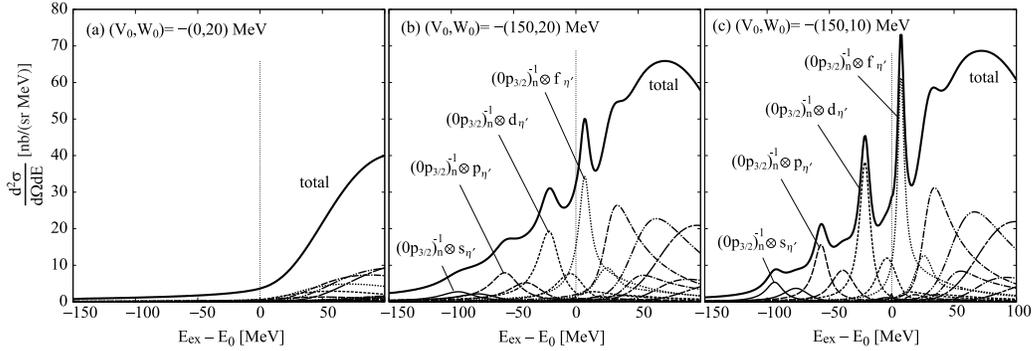}
\caption{Calculated spectra of the $^{12}$C$\reaction^{11}$C$\otimes\eta'$
 reaction at $T_p=2.50$ GeV as functions of the excitation energy $E_{\rm
 ex}$.  $E_0$ is the $\eta'$ production threshold energy.  The
 $\eta'$-nucleus potentials are (a) $(V_0,W_0) = -(0,20)$ MeV, (b)
 $-(150,20)$ MeV, and (c) $-(150,10)$ MeV.  The total spectra are shown
 by the thick lines, and the dominant configurations are also shown.
 The neutron-hole states are indicated as $(n\ell_j)_n^{-1}$ and the
 $\eta'$ states as $\ell_{\eta'}$.
}
\label{fig:theoertical_spectra} 
\end{center}
\end{figure}

In Fig.~\ref{fig:theoertical_spectra}, we show the calculated spectra of
$\eta'$ mesic nuclei formation for the $\reaction$ reaction on $^{12}$C
for three different sets of potential depths
as functions of the excitation energy $E_{\rm ex}$ near the
$\eta'$ threshold $E_0$.
The kinetic energy of the incident proton is set to be 2.50
GeV. The spectrum in Fig.~\ref{fig:theoertical_spectra}(a) is obtained
with the potential parameter $(V_0,W_0)=-(0,20)$ MeV which means that
the $\eta'$ meson mass does not change in the medium.  In this case, we
do not see any bound states in the spectrum because there is no
attraction, and we only see the quasi-free $\eta'$ contribution.  In
Figs.~\ref{fig:theoertical_spectra}(b) and (c), we show the spectra
with the mass reduction of the $\eta'$ meson.  We put $V_0=-150$ MeV
corresponding to the 150 MeV mass reduction at the normal saturation
density~\cite{Costa:2002gk,Nagahiro:2006dr}.  The strength of the
imaginary part is set to be (b) $W_0=-20$
MeV~\cite{Nagahiro:2006dr} and (c) $-10$ MeV. 
The data of the transparency ratio measurements~\cite{Nanova_PLB,Nanova_PPNP} suggest that 
the absorption width is
15--25 MeV, which corresponds to $W_0 \simeq 
-(7.5\mbox{--}12.5)\,\mathrm{MeV}$. 
Because
the momentum transfer for the $\eta'$ production is not small, many
components with higher angular momentum give finite contributions,
and therefore the resulting spectra are not simple.  We observe several
peak structures in the bound region $E_{\rm ex}-E_0 < 0$ owing to the
bound states.  A relatively large peak is seen around the threshold
$E_{\rm ex}-E_0 \sim 0 $ MeV originating from the so-called threshold
enhancement of the quasi-free $\eta'$ production. 
These peak structures are a signature of the attractive potential.


\section{Experimental Concept}
\subsection{Inclusive measurement}
Let us assume here that the decay width of $\eta'$ mesic nuclei is small
compared to the binding energy.
Then, an inclusive measurement, detecting the ejectile
but not the decay particle, will work
to search for a possible signal of $\eta'$ mesic nuclei.
The appearance of a statistically significant peak
requires not only high statistics but also a good resolution
because of a huge background.
Those requirements will be discussed and the fulfillment will be proven
in the following sections for a realistic case at the existing facilities of GSI-SIS.

An inclusive measurement has an advantage
over an exclusive measurement, in which the decay particles will also be detected.
The detection of a particular decay particle distorts the missing-mass spectrum
because its emission probability and its detection efficiency have to be folded in,
while an inclusive measurement is not biased by properties of the decay process
even though the signal-to-noise ratio is worse.
The comparison of these two kinds of measurements may yield further information on 
formation and decay processes of the $\eta'$ mesic nuclei.

\subsection{Experimental principles and setup}
\begin{figure}[t]
\begin{center}
\includegraphics[angle=0, width=0.90\textwidth]{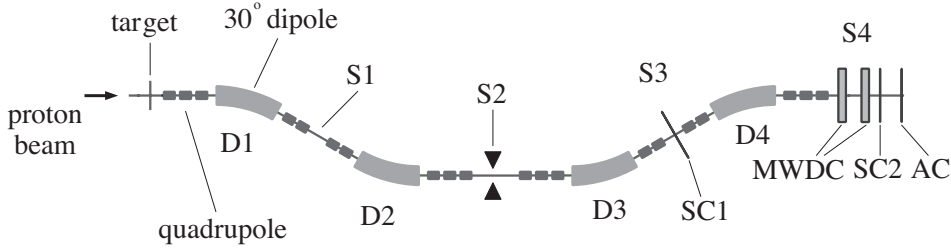}
\caption{Schematic view of the FRS. Proton beam of $T_p = 2.50$\ GeV is extracted
from SIS and transferred to the target. The outgoing deuteron from the
$\reaction$ reaction makes a target image on the central focal plane at S2.
We place slits at S2 to collimate the image. The momentum of the deuteron 
is analyzed and measured by MWDCs located at the final
focal plane at S4.
Segmented scintillation counters at S3 (SC1) and S4 (SC2), and Cerenkov counter at S4 (AC) will be used for particle identification and for the hardware trigger of the data acquisition. 
}
\label{Fig: FRS}
\end{center}
\end{figure}

Here we investigate the experimental feasibility of the inclusive spectroscopy 
to be performed at the GSI-SIS facility.
The incident proton beam with the energy of $T_p = 2.50$\ GeV will be extracted from 
the SIS, and the fragment separator FRS~\cite{FRS} will be used as a spectrometer
as depicted in Fig.~\ref{Fig: FRS}.
We install our target materials, carbon and carbon deuteride, 
in the nominal target position of the FRS and measure the 
excitation spectra for both the $^{12}{\rm C}\reaction$ and 
the ${\rm D}\reaction$ reactions. 
The spectrum with the deuteron target,
which is expected to be smooth,
will provide a reference to the quasi-free background for $^{12}\mathrm{C}$ target
(discussed in Section 4.1).

The proton beam intensity will be $10^{10}$/spill 
(spill length of 1 second, 6 second-cycle).
The target will have a shape of 2 mm-wide strip with
the thickness of $\sim$ 4 g/cm$^2$.
We can adopt the thick target without seriously deteriorating the
spectral resolution since the incident beam and the 
outgoing particle have similar energy loss in the target. 
This will lead to an experimental advantage of larger luminosity. 

The ejectile deuteron with the momentum of $\sim 2.8\ \mathrm{GeV}/c$
is analyzed by the FRS from the target to the 
final focal plane S4 with the resolving power of about 2000.
Two sets of conventional-type multi-wire drift chambers (MWDCs) will be installed near the
S4 focal plane and measure the deuteron tracks. 
MWDCs will be placed with 1 meter distance between each other
and the overall horizontal tracking resolution 
will be $\sim$ 150 $\mu m$ (FWHM). 

For particle identification purposes, a segmented scintillation 
counter will be installed at S3, and another segmented scintillation counter and 
a $\Cerenkov$ counter at S4. Particles are identified by 
Time-Of-Flight (TOF) measured in the S3-S4 section in the software analysis.
The typical TOF resolution of 0.3 ns ($\sigma$) is sufficient for the discrimination of 
the signal deuterons from the background protons by the TOF difference of about 10 ns.

A key of the experimental feasibility is the background, instrumental one and
physical one. For the instrumental background, a number of particles are expected to emerge from
the intense primary beam hitting the beam pipe near the S1.
The S1-S3 section will be used to sweep out these particles and to select
particles originating in the production target by inserting slits at the
S2 position.

The physical background level of protons from the target is estimated to be of the order of
500 $\mu {\rm b/sr/(MeV}/c)$ by using the FLUKA package~\cite{FLUKA}.
It corresponds to $\sim 6 \times 10^{4}$/spill.
Thus, we need to suppress the background at the hardware level by 
an order of 100 to comply with the DAQ speed.

In combination with the segmented scintillation counters, 
a velocity sensitive $\Cerenkov$ counter needs to be developed and installed at S4 to provide
the hardware-level trigger.
The deuteron central velocity $\beta$ is 0.83 while the background protons have $\beta = 0.95$.
The $\Cerenkov$ counter consists of aerogel\footnote{
High refractive index aerogel radiator is developed in Chiba University~\cite{Adachi}.}
(refractive index $n=1.12$) and 
will provide a veto for the background protons.
According to a test experiment at Tohoku University,
more than 95 \% of protons will be eliminated
by setting the threshold to two photons~\cite{LNS2008}.

\begin{table}[t]
\caption{Contributions to the spectral resolution. }
\label{Table: resol}
\begin{center}
\begin{tabular}{cc}
\hline
\hline
Factor & Contribution in $\sigma$ [MeV]\\
\hline
\hline
Beam momentum spread & 0.40 \\
Energy loss in target & 1.3  \\
Spectrometer & 0.77 \\ 
\hline
Total & 1.6 \\
\hline
\hline
\end{tabular}
\end{center}
\end{table}

The overall spectral resolution is estimated to be 1.6\ MeV ($\sigma$), which is
sufficiently smaller than the natural widths of the bound states.
Table~\ref{Table: resol} lists each factor contributing to the resolution.
We also measure elastic D$\reaction$ reactions~\cite{Berthet82}
to evaluate directly the spectral resolution.

\section{Simulation of Inclusive $\reaction$ Spectrum}
In order to investigate the feasibility of the inclusive spectroscopy at GSI-SIS,
a simulation was carried out.
Both the $\eta'$ production and other background contributions are taken into account.
As stressed in Section 2, the most prominent peak structure is expected to appear around the threshold. Thus we concentrate on the possibility to observe this structure, as a signature of
an attractive $\eta'$-nucleus interaction or the mass reduction of an in-medium $\eta'$.

\subsection{Background cross section}
The dominant contribution of energetic
deuterons comes from quasi-free processes, such as $pp\to dX$ and
$pn\to dX$ ($X=2\pi,\ 3\pi,\ 4\pi,\ \omega$). In evaluating each cross
section, we adopt almost the same approach as the background study for
the measurement of the $pn\to d\omega$ reaction at
COSY-ANKE~\cite{Barsov04}, where the beam energy is similar to our
case. The total cross section of multi-pion production can be
parametrized as a function of the center-of-mass energy, and the same
parameter sets as described in Ref.~\citen{Barsov04} are used. As for the
$pn\to d\omega$ contribution, the near-threshold cross section data
obtained at COSY-ANKE~\cite{Barsov04} are extrapolated to the energy
region of our interest.
The uncertainty of these cross sections
is not taken into account.
It is worthwhile to note that a simulation
with the same parameters reproduces the differential cross section at
$0^\circ$ for the $pp\to dX$ reaction at
$T_p=2.5\ \mathrm{GeV}$~\cite{Turkot63} fairly well.

\begin{wrapfigure}{2}{6.6cm}
\includegraphics[width=6.0cm]{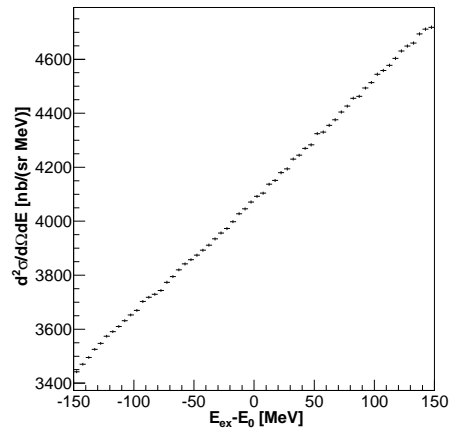}
\caption{Estimated background cross section of the $p+^{12}\mathrm{C}\to d+X$ reaction as a function of the excitation
energy. $E_\mathrm{ex}$ denotes the excitation energy
and $E_0$ does the $\eta'$ emission threshold energy.
The error originates from the statistical error in the simulation,
and does not mean the uncertainty of the cross section.
}\label{p_d_reaction_carbon_ex}
\end{wrapfigure}

The result is shown in Fig.~\ref{p_d_reaction_carbon_ex}.
The effective nucleon number for the $^{12}\mathrm{C}(p,d)$ reaction
is theoretically estimated to be 1.1 for each quasi-free process,
based on the eikonal approximation~\cite{Hirenzaki92}.
The theoretical predictions and the background estimations
are considered to include various ambiguities.  Thus, to perform the 
simulation of the experimental spectra,
we adopted here the severer conditions by considering the effective nucleon number to be two. 
The background level ($\approx \mathrm{\mu b/(sr\, MeV)}$) is found
to be two orders of magnitude larger than
the peak from $\eta'$ mesic nuclei in Fig.~\ref{fig:theoertical_spectra}.

\subsection{Expected spectrum and experimental feasibility}
\begin{figure}[t]
\begin{center}\includegraphics[angle=0, width=\textwidth]{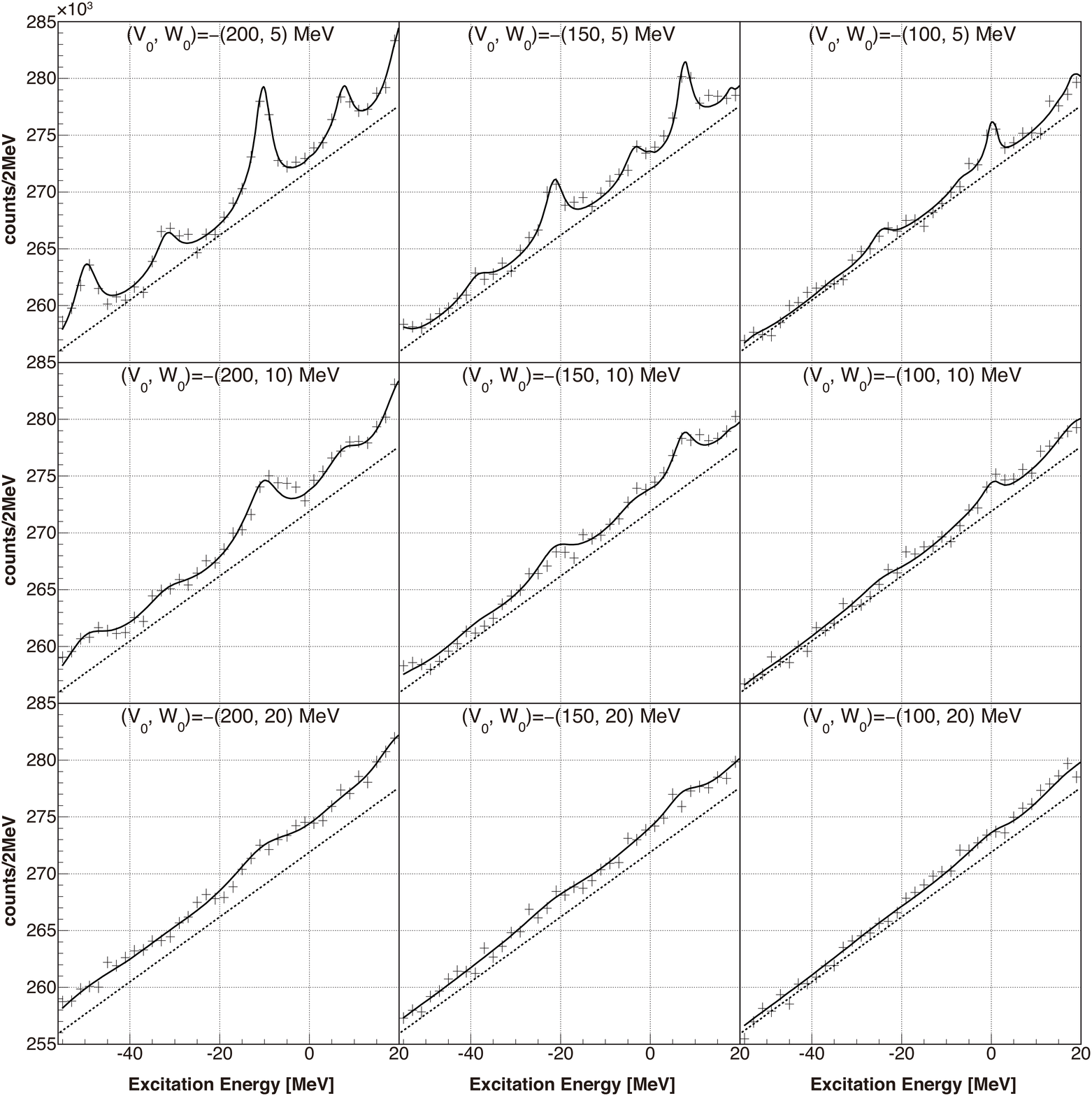}
\caption{Simulated spectra with $3.24\times 10^{14}$ protons on a $4\ \mathrm{g/cm^2}$-thick $^{12}\mathrm{C}$ target. The dashed line corresponds to the background processes discussed in Section 4.1, and the solid line includes the contribution from the signal process discussed in Section 2.2.}\label{simulated_spectra}
\end{center}
\end{figure}


Figure~\ref{simulated_spectra} shows the result of the simulation for $V_0=-200$, $-150$,
$-100\ \mathrm{MeV}$ and $W_0=-5$, $-10$, $-20\ \mathrm{MeV}$, which corresponds to $3.24\times 10^{14}$ protons on a $4\ \mathrm{g/cm^2}$-thick $^{12}\mathrm{C}$ target\footnote{Requested for each FRS mode in the Letter of Intent for GSI-SIS~\cite{LoI}.}.
If the real part of the potential ($-V_0$), namely the mass shift of an $\eta'$ meson at the nuclear saturation density, is sufficiently large and the imaginary part ($-W_0$) is less than $10\ \mathrm{MeV}$,
the near-threshold peaks will be observed with enough statistical significance
against a poor signal-to-noise ratio of the order of 1/100.
However, a smaller mass reduction and/or a larger absorption makes the signal-to-noise ratio even worse, and experimental observation would be more difficult then.

\begin{wrapfigure}{1}{6.6cm}
\includegraphics[width=6.0cm]{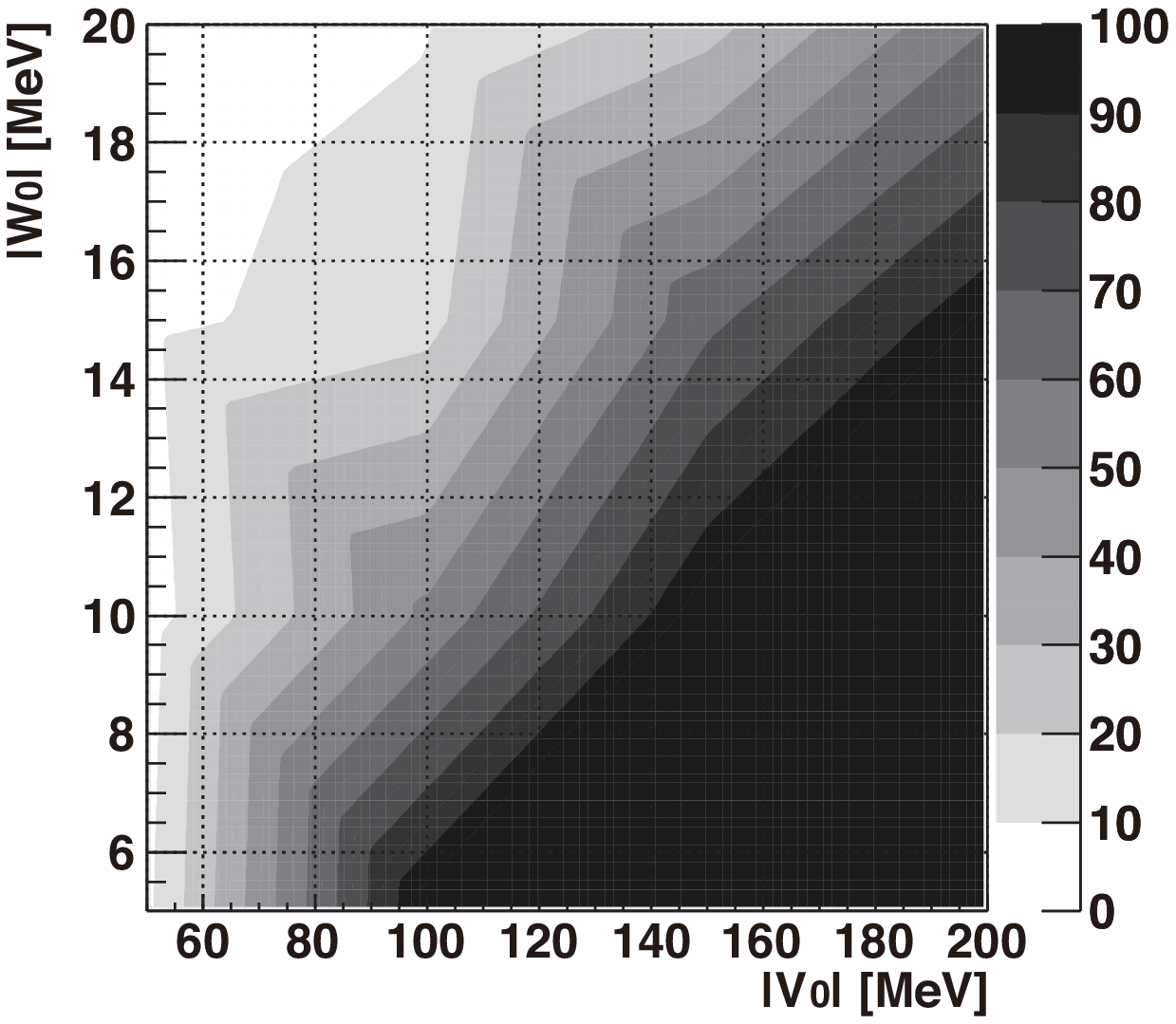}
\caption{The probability (in unit of \%) of rejecting the null hypothesis at 95\% C.L., as a function of the optical potential ($V_0+iW_0$).}\label{chi2}
\end{wrapfigure}

Furthermore, we evaluated the sensitivity of observing the peaks for each optical potential
by a $\chi^2$ test for the null hypothesis that there were only a structureless background\footnote{Each spectrum is fitted by a third-order polynomial function.} in the range of $-56\ \mathrm{MeV}<B<20\ \mathrm{MeV}$.
If the $p$-value, defined as the probability to get a $\chi^2$ greater than the computed $\chi^2$, is less than a certain value (set at 0.05 here), we will reject the null hypothesis. Figure~\ref{chi2} shows the probability to ``find'' a statistically significant ($95\%$ C.L.) structure as a function of the optical potential parameters.

If the strength of the attraction at the saturation density ($-V_0$) is as large as $150\ \mathrm{MeV}$ predicted by the NJL model~\cite{Costa:2002gk,Nagahiro:2006dr} and the absorption ($-W_0$) is less than $12.5\ \mathrm{MeV}$ as indicated by the CBELSA/TAPS experiment~\cite{Nanova_PLB,Nanova_PPNP}, we will have a large chance to observe peak structures related to $\eta'$ mesic nuclei, given a reasonable number of protons as beam.
\section{Summary}
We have discussed an inclusive measurement of the $\reaction$ reaction for investigating $\eta'$ mesic nuclei, and have shown that it may be feasible at GSI-SIS with the FRS. The signal to noise ratio is estimated to be of the order of 1/100, when the experimental resolution by use of the FRS is negligibly small ($\sigma \sim 1.6\ \mathrm{MeV}/c^2$) compared to the decay width of $\eta'$ mesic nuclei. We conclude that the observation of the enhanced peak is
reasonably achievable even with such a poor signal strength, 
by accumulating an immense number of events,
together with an intense proton beam supplied by the SIS synchrotron.
\section*{Acknowledgements}
This work is supported by the Grant-in-Aid 
for Scientific Research
(No. 20002003, No. 20540273, No. 22105510, No. 22740161, No. 24105705,
No. 24105707, No. 24105712, and No. 24540274) in Japan.
Financial support by HIC-for-FAIR is highly appreciated.
This work was done in part under the Yukawa International Program for Quark-hadron Sciences (YIPQS).
\appendix
\section*{Cross Section of Elementary Process}
Since there is no experimental
information on the elementary process $pn\to d\eta'$ itself at this moment,
its total cross section at $T_p=2.50\ \mathrm{GeV}$ has been evaluated in
the following two ways.

According to the prescription described in Ref.~\citen{Rejdych07}, the ratios
of the cross sections
$\sigma(pn\to d\eta')/\sigma(pp\to pp\eta')$
and
$\sigma(pn\to d\eta)/\sigma(pp\to pp\eta)$
will be of the same order. 
The measurements at CELSIUS/WASA revealed the cross section for
$pn\to d\eta$ is one order of magnitude larger than that for $pp\to
pp\eta$ at the excess energy around
$30\mbox{--}40\ \mathrm{MeV}$~\cite{Calen97}, which is the region of
our interest. By using the cross section for the $pp\to pp\eta'$ ($\sim
0.2\ \mathrm{\mu b}$) obtained by the COSY-11
collaboration~\cite{Khoukaz04}, the cross section for 
the $pn\to d\eta'$ reaction will be around
$3\ \mathrm{\mu b}$.

Grishina {\it et al.} performed a two-step model calculation for
$pn\to d\eta$ and $pn\to d\eta'$ reactions~\cite{Grishina00}. The
experimental data of the $pn\to d\eta$ cross section is reproduced by
taking into account not only $\pi$ exchange but also $\rho$ and
$\omega$ exchange, and the effect of initial state interaction. A
similar calculation for $\eta'$ shows the cross section for $pn\to
d\eta'$ is $\sim 3\ \mathrm{\mu b}$ with an uncertainty of factor 1.5,
due to the S-wave $\rho^0 p\to \eta' p$ amplitude.

By assuming an isotropic distribution for the $pn\to d\eta'$ reaction, the
differential cross section at zero degree in the laboratory frame is
estimated to be $\sim 30\ \mathrm{\mu b/sr}$.

Finally, we would like to make a few comments on the uncertainties of the elementary cross section.  
We have evaluated the elementary cross section of $pn\to d\eta'$ production
by using the experimental values as discussed above.  
However, there will be some uncertainties for the elementary cross section
due to the energy and momentum dependence of the 
elementary process for the $\eta'$ meson production in a nucleus as considered in this article. 
The experimental information on the elementary $\eta'$ production process 
is rather poor and it is very difficult to know the momentum and energy dependence of the elementary cross section.  
Thus, we discuss the dependence using the data of the $\eta$ (instead of $\eta'$)
meson production process, 
which are also used partly to evaluated the $\eta'$ production cross section.  

In Refs.~\citen{Calen97,Calen98}, the data of $pn\to d\eta$ cross section are reported around the production threshold up to the energy 
corresponding to the $\eta$ excess energy $Q= 120\ \mathrm{MeV}$.
For the $\eta'$ meson production considered in the article, 
the corresponding region of the $\eta'$ excess energy $Q$ is
$Q=20 \mbox{--} 50\ \mathrm{MeV}$ by considering the $\eta'$ fermi motion in nucleus.  
In the region, the data of the $\eta$ meson production show the 
smooth and monotonic increase of about a factor 3 as a function of $Q$. 
Thus, we expect that the actual value of the elementary cross section 
will be the average of the cross section in this region and that the momentum and energy dependence of the elementary cross section will not provide 
extra structures in the spectrum except for a smooth slope.
The size of the whole calculated signal spectrum in the nuclear target reaction is proportional to 
the size of the elementary cross section. 
Thus, we think the ambiguities of the size of the elementary cross section 
will affect the size of the whole $(p,d)$ spectra but will not affect the peak structure of the spectrum much.  
This expectation will be correct especially for the cases with relatively narrow peaks such as considered in this article.


\end{document}